# Gravity: a blockchain-agnostic cross-chain communication and data oracles protocol

*A Preprint*

Aleksei Pupyshev, Dmitry Gubanov, Elshan Dzhafarov,
Ilya Sapranidi, Inal Kardanov, Vladimir Zhuravlev, Shamil Khalilov,
Marc Jansen, Sten Laureyssens, Igor Pavlov, Sasha Ivanov

Gravity Team

*oracle@gravity.tech*

## Abstract

This paper intends to propose the architecture of a blockchain-agnostic protocol designed for communication of blockchains amongst each other (i.e. cross-chain), and for blockchains with the outside world (i.e. data oracles).

The expansive growth of cutting-edge technology in the blockchain industry outlines the need and opportunity for addressing oracle consensus in a manner both 'technologically and economically' efficient as well as futureproof. Blockchain-agnosticism is inherently limited if proposing a technological solution involves adding one more architectural layer. As such, Gravity protocol is designed to be a truly blockchain-agnostic protocol. By ensuring parity through direct integration and by leveraging the stability and security of the respective interconnected ecosystems, Gravity circumvents the need for a dedicated, public blockchain and a native token. Ultimately, Gravity protocol intends to address scalability challenges by providing a solid infrastructure for the creation of gateways, cross-chain applications, and sidechains. This paper introduces and defines the concept of Oracle Consensus and its implementation in the Gravity protocol named the Pulse Consensus algorithm. The proposed consensus architecture allows Gravity to be considered a singular decentralized blockchain-agnostic oracle.

# Introduction

Blockchain technology addresses technical challenges associated with centralized systems, by proposing an inherently radical approach to immutability and efficiency through decentralization. This core ideology has seen such drastic innovation that consequentially the entire industry is rapidly revolutionizing all aspects of communication protocols and open-source ecosystems. Through cryptocurrencies, smart contracts and cross-chain communication infrastructures, cutting-edge financial frameworks and instruments are emerging, and lie at the intersection of economics, law, sociology, mathematics and computer science, which are generally referred to as Decentralized Finances (DeFi) or Web 3.0.

For the World Wide Web to initially expand so extraordinarily, it had to confront the need for instant and reliable exchange of information, through flexible and scalable communication networks. The growth of the Internet was accompanied by an increasing level of network decentralization, autonomy, and automation, made possible by experimentation with and development of various protocols for data transfer between network hubs. This need for experimentation and innovation, especially regarding cross-chain communication, is still present today in the blockchain ecosystem, in order to achieve what ultimately is our common goal, a global, interconnected decentralized architecture accessible to everyone and capable of driving adoption to new revolutionary ideas beyond what is possible with Web 2.0.

One of the final required technological solutions is the capacity to transfer data from real-world information sources to a network of blockchains and to establish communication between these independent networks. To accomplish this, one has to rely on so-called oracles. However, paradoxically, one can conclude that in such a trustless system, there remains a need for trust in individual elements. It is of crucial importance for the entire industry to research and develop new frameworks in order to solve trustless communication.

A significant distinction is observed between existing solutions and what ideologically, an efficient solution could look like. A common design of currently proposed architectures is to solve blockchain-agnostic communication by interjecting an additional layer to the protocol, and thus, an extra layer of complexity. One cannot but acknowledge that technology and finance are revolutionized and improved upon by blockchain, yet this seems to go hand in hand with the ideology that most, if not all, blockchain-related innovation requires a proprietary chain and token. This paper introduces a solution to solve Oracle Consensus, stripped away from a proprietary blockchain and/or token, with the intention to achieve increased efficiency, security and inclusivity. Gravity, by its very nature, is a proposal for a truly blockchain-agnostic protocol.

Eliminating the requirement of forming a new blockchain to solve cross-chain interaction, remarkably impacts the technological and economical outcome of the proposed protocol. In the context of Gravity, cross-chain communication is when network A communicates directly with network C, without interjecting a new proprietary network B. This results in a more efficient solution by removing that layer of complexity. However, more importantly, having a proprietary token for a blockchain-agnostic protocol interferes with the incentive alignment of the internal ecosystem. Blockchain-agnosticism can theoretically not be achieved if the proposed solution allows for the potential of a future rival clash with one, if not all, the chains it connects. Any protocol eliminating such a tribalistic design limitation can enable synergies between ecosystems that otherwise would compete. Additionally, Gravity's internal economy leverages the native economies of the chains it connects. Compared to underwriting the intrinsic, economical security in a consolidated financial structure, Gravity underwrites its economic security through the cumulative decentralized stability of all Gravity-connected chains.

For illustrative purposes, the significant differences between Gravity and the most popular current solutions for oracles and interoperability are presented in Table 1:

|  | **Gravity** | **ChainLink** | **Polka** |
|---|---|---|---|
| Incentivization | TOKENS OF INTEGRATED CHAINS | DEDICATED TOKEN | DEDICATED TOKEN |
| Architecture | INTERNAL LEDGER | DEDICATED TOKEN | DEDICATED BLOCKCHAIN & TOKEN |
| Network administration | DECENTRALIZED | CENTRALIZED | DECENTRALIZED |
| Network governance | DECENTRALIZED | CENTRALIZED | DECENTRALIZED |
| Dedicated consensus between nodes | YES (PULSE) | NO | YES (GRANDPA/BABE) |
| Network entry for new nodes | OPEN | CLOSED | OPEN |
| Blockchain-agnostic* | YES | NO | NO |
| Blockchain interoperability solution | YES | NO | YES |
| Data oracles solution | YES | YES | NO |
| Sidechains solution | YES | NO | YES |

*According to the Statement of Principles.

## Statement of Principles

1. To advance the industry of decentralization and open finance, we need true blockchain-agnostic solutions for oracles and interchain communication.
2. A solution cannot be fully blockchain-agnostic if it uses a dedicated blockchain with its own native token.
3. A dedicated token complicates the interaction between oracles and should not be necessary to pay for oracle services.
4. Oracle consensus is a truly decentralized innovation, whereas any alternative solutions are essentially groups of centralized oracles.

5. The blockchain industry needs a unified, trusted, decentralized blockchain-agnostic oracle, which is an open-entry network rather than a marketplace of individual independent oracles.
6. The economy of the unified blockchain-agnostic oracle should be supported by the native token economies it connects and, by consequence, could positively impact their growth.
7. Creating the unified decentralized oracle is a common global challenge for all of the public blockchain ecosystems to ensure their symbiosis and prosperity.

## Current Challenges

In this section, we describe a tentative list of issues specifically related to current cross-chain communication and data oracle systems, aimed at reviewing a broad set of obstacles not necessarily solved in its entirety by the Gravity protocol. This section provides a useful contextual background to facilitate the explanation of concepts proposed in this paper.

**Challenge I**: Cross-chain communication is unachievable in the absence of trusted oracles or validators.

Due to the importance of Challenge I, the focus of this paper lies mainly on a detailed description of the protocol for the transfer of data from external sources into target blockchains. In that context, one blockchain serves as an external data source for another blockchain.

**Challenge II**: The use of several redundant primary data sources is needed to improve the reliability and accuracy of the aggregated data.

One isolated external data source can – accidentally or maliciously – behave in a way that could be dangerous to the security of the system, either by reporting incorrect information or due to being subjected to censorship, blocking or filtering data, or any other external interference. It is the diversification of data sources that can help to increase the resilience of a system to the described threats.

**Challenge III**: A single oracle is less trustworthy than a consortium of several independent oracles, where each oracle has the same level of trust in the system.

This problem can be demonstrated by calculating probabilities, but it is also intuitively evident that several votes from independent organizations are more reliable than a vote from one with the same level of trust. This issue can be addressed by integrating a threshold signature (or a multi-signature), where data is verified simultaneously by multiple participants, which decreases the probability of a successful attack.

**Challenge IV**: Oracle systems require trust.

Trust is a fundamental concept that represents the level of risk to individual users of a system, where the higher the trust, the lower the risk assessment of the extent of potential damage in the event of force majeure or attack is. The trust in a system is essentially the sum total of trust in individual oracles, which is dependent on the history of actions inside and outside the system and on internal mechanisms that conduct monitoring and evaluation of the reputation of oracles.

To maintain the trustworthiness of oracle networks, it is of absolute necessity to incorporate built-in security systems and self-governance strategies, capable of working automatically and for manual interventions. For instance, an oracle that has become malicious and capable of affecting the entire system, should automatically be excluded from the consensus of active oracles that sign the data and deliver it to a target blockchain.

**Challenge V**: A certain level of commitment is needed on the part of data providers In a trust-based system, as the role of an oracle is critical and significant.

Depositing tokens for a relatively long period of time can serve as confirmation of "goodwill" and is the first indicator of how trustworthy a new oracle is.

**Challenge VI**: The improper performance or malicious activity of an oracle can negatively influence the outcome of data provision, unless it is prevented by imposing financial consequences for its owners.

There are at least two ways of imposing sanctions on an oracle that suddenly becomes malicious: "slashing" (a penalty that occurs under certain conditions), or freezing the deposited funds for an extended period of time. In this paper, we only consider the latter.

**Challenge VII**: Distributed systems are more prone to attack, if attempts to do so are less expensive than expected gains and are not excessively time consuming.

Delaying the deposit lock for an extended period of time, after the oracle voluntarily leaves the system, is a security measure aimed at introducing complications for attacks such as the Sybil attack, which consists of registration and re-registration of a large number of nodes.

**Challenge VIII**: Data verified by oracles requires a collectively verifiable on-chain signature (threshold signature) in order to be securely delivered to a target blockchain.

Otherwise, independently of what result has been obtained within the internal consensus of the oracles, the entity that signs the data in the target blockchain would remain subject to an attack.

**Challenge IX**: Proof that the data was obtained in an original way by each oracle independently of the others is needed to allow for oracle monetisation.

To protect against the problems represented by Challenge X, oracles need to request the information from external sources independently, rather than borrow the transmitted data values from each other. Such problems are solved by the scheme of commit-reveal, where at the onset, only the hash generated from the data is revealed, with the values disclosed afterwards. If the disclosed hash does not correspond to the disclosed data, it is recognised that there has been an attempt of fraudulent data provision.

**Challenge X**: In order to monetise the work of oracles, verification of obtained data and its delivery to user smart contracts must be separated, and the data itself must be disclosed (decrypted) only at the time of delivery to the client.

The process of verifying the data without disclosing it in a target blockchain allows the data to be kept secret and disclosed only when delivered to the recipient, who has the right to determine its availability to other smart contracts in the blockchain. This solution increases the incentive for oracles to deliver data by safeguarding ways to receive rewards for data provision.

## Key Concepts

**Oracle consensus** is an agreement in the oracle network about which data feed value to consider valid under certain conditions. Within Gravity, oracle consensus is achieved by on-chain verification of the decision made by the oracle system in the target blockchain. Unlike block generation consensus mechanisms, oracle consensus deals with a more complex system composed of different data streams and applications that utilise them.

**Pulse consensus** is an implementation of oracle consensus in the Gravity protocol. It consists of two stages: Commit-Reveal & Data Aggregation and Multi-Signature On-Chain Verification. The mechanism of selecting a subset of oracles to participate in data verification is contingent upon each node's reputation.

**Node** is a "building-block" element of the Gravity network. It is a middleware that owns and operates accounts in all supported target blockchains and in the internal distributed ledger. Each node is composed of infrastructure components such as data extractors, task scheduler, and connectors to target blockchains and the internal ledger.

**Data feed** is a specification-defined data source. An example of a data feed would be the number of newly registered COVID-19 cases in the world per day.

**Raw data** is unprocessed data from primary sources (external data APIs).

**Data** is output data processed by a node's extractor. Extractors allow for the use of certain operations on data points collected from different sources and numerical transformations into a format suited for the specific purpose and the specific target blockchain. For example, extractors can convert decimal data values into integers.

**Agg. data** is aggregated data from all nodes involved in data delivery with a certain numerical transformation applied to it (for example, an average, median, or the most frequent value).

**Target blockchain** is a supported blockchain network where data is written into by nodes. It contains smart contracts to verify signatures and smart contracts of users who pay for data delivery.

**Oracle** is an account in a target blockchain that signs and supplies data.

**Proof** is a cryptographic signature that takes messages and seeds for generation. It is verified within the smart contract and is matched and compared with the corresponding public key of the oracle.

**USER-SC** is a user application (smart contract) in a target blockchain, which receives data from the Gravity system via a subscription model.

**NEBULA-SC** is a smart contract in the target blockchain used by a number of oracles that supply a data feed under certain conditions (price for delivery, reputation threshold, minimum required number of oracles). This contract verifies the threshold signature parameters, accumulates payments from users and controls the distribution of rewards among the oracle providers of the nebula.

**SYSTEM-SC** is the main register of information on active Gravity nodes and their reputation scores. Each target chain has its own instance of the system smart contract. It manages both the registration of new nodes, and the deposit transactions for the nodes that have chosen the respective target chain for conducting their operations. It also serves as the register for all supported data feeds and nebulae in the corresponding target chain.

**Internal Distributed Ledger (IDL)** is a "software message bus" that supports communication of Gravity nodes with each other and provides storage with a quick finalization consensus (e.g.: BFT).

**Consuls** are nodes with the highest reputation score in the network that acquire special functions within the system. They are authorized to update/migrate the system smart contracts, nebula smart contracts and serve as consensus validators in the internal ledger.

**Pulsation** is the continuous data delivery process that generates pulses. Each pulse starts with a task to deliver data feeds to a target blockchain and results in successful verification and delivery of data to the subscribers (USER-SC).

**Leader** is a node selected to initiate data transfer transactions in the target chain within the current pulse. It invokes two types of transactions in the target chain: pulseTx, which verifies the hash generated via multi-signature from data aggregated within the Gravity system, and sendDataTx, which delivers the verified data to USER-SC. In addition, the leader collects hashes and proofs of aggregated data from all nodes that provide data feeds. The leader selection is based on the rules described in NEBULA-SC (for instance, nodes can alternate as leaders depending on the height of the target chain).

**Subscription** is a user's indication of the user contract and the public method that should be used to receive verified aggregated data.

**Reputation** is a numerical representation of the confidence level put into a Gravity node from its peer nodes, calculated on the basis of regular mutual evaluations of all nodes.

## Data Provision Workflow: Pulsation

Typically in oracle systems, a common workflow pattern is used, which contains three global entities: 1. the external world with data sources, 2. the oracle system, and 3. the target blockchain where data should be safely recorded. Figure 1 shows the scheme of data provision workflow in the Gravity system, which is called pulsation. Data delivery started at a certain point in time (block or a block interval) should consistently end with the verification of the pulse transaction. If the verification fails for some reason, data delivery for the current time point is not performed. A detailed description of the stages of data delivery from the outside world to the target-chain can be found in chapters Commit-Reveal & Data Aggregation Flow and Threshold Signature & Data Verification Flow.

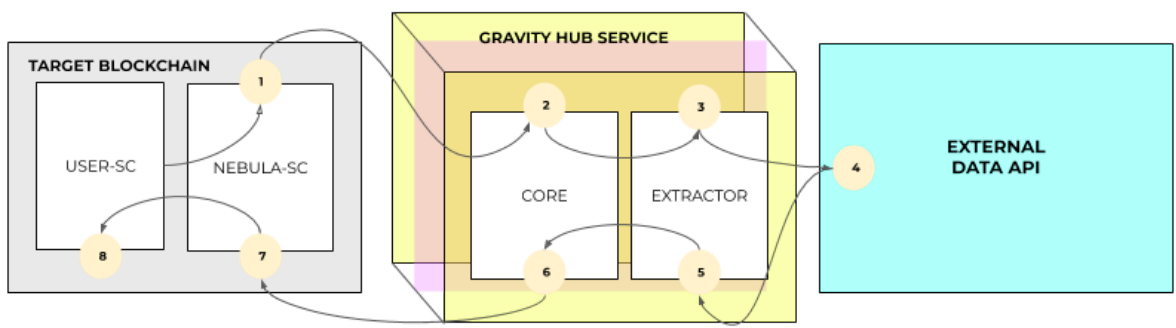

Figure 1: Data provision workflow (pulsation):

1) The user chooses a data feed and the contract of a nebula that they want to subscribe to;

2) The core of the node monitors the status of data delivery and requests data from the data extractor;

3) The extractor accesses external data sources in an asynchronous manner, independent of the target blockchains;

4) The requested raw data is fed into the extractor;

5) The extractor processes the data, performing all necessary aggregation and filtering;

6) Each Gravity Node's core initiates and conducts the commit-reveal process, including data aggregation between peers and collecting multi-signatures from peers;

7) The calculated hash from the data aggregated by the oracle (leader) is delivered to the nebula contract in the target chain, where the signatures and other necessary conditions are verified (pulse tx);

8) The verified data is delivered to the USER-SC as a subscription service.

Becoming a part of the data provision workflow, the user is not obligated to choose a specific oracle or their subset to trust. Instead, the entire data provision service of the Gravity network provides all necessary checks within itself to be ultimately trustful to the final user, independent of their requirements. In addition, the user is not required to apply their own methods of data aggregation, as it can be executed automatically within the system, which can provide finalized and ready-to-use numbers or string values from a data feed to user applications (USER-SC).

The security verification of the transferred data is carried out within NEBULA-SC deployed in target blockchains. NEBULA-SC contains parameters and instructions for verification of signatures of the participating data providers, such as: public keys of active providers, threshold signature settings, the reputation threshold for node admission and the leader selection rule. The Pulse Leader is a node that collects all signatures and aggregated data from peers and sends them for verification to the NEBULA-SC contract by calling the pulseTx method.

The number of potential NEBULA-SC contracts in the Gravity network is unlimited. Each user or node operator can create a contract with custom properties by paying a fee in one of the tokens supported by the hostchain.

# Architecture

## *Node Structure*

The key element of the Gravity network is the Gravity node. On the whole, nodes are non-isomorphic, meaning that providers can freely choose to work in one or several target chains, or to implement extractors not for all possible data feeds, but only for a relevant subset. The main components of a Gravity node are the core, responsible for all business logic implementing the protocol, and so-called data feed extractors. In Figure 2, the overall structure of a node is described.

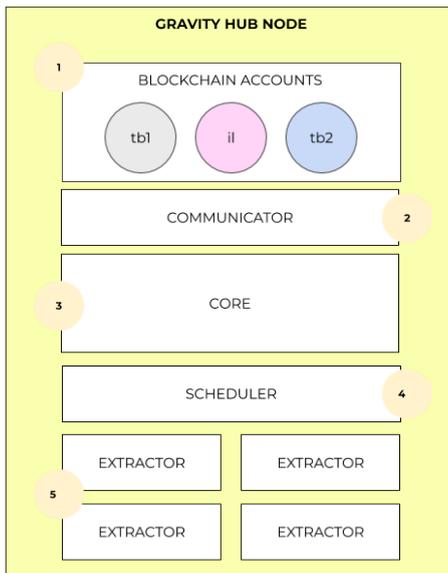

Figure 2: Node structure

1) Each node has a public key in each of the supported target blockchains and in the internal ledger;

2) The communicator module is used to establish communication with one of the supported target blockchains and the internal ledger;

3) The key logic of the Gravity node for reputation calculations, commit-reveal operations, aggregation and coordination of the rest of the components is implemented in the core module;

4) The Scheduler manages processes that depend on the time and status of the tasks. For example, it can start scheduled data delivery to target chains and extraction from external sources based on a certain time condition;

5) Extractors are services that request and process data feeds from external sources. One extractor corresponds to one data feed supported by the node but can use and combine multiple data sources.

Datafeeds supported by a Gravity node can be described as a boilerplate source code or as an implementation in the form of a data extractor. Each extractor collects data in accordance with the specification for the required data. The specification defines:
- where to get data from (recommended sources or mandatory sources)
- how to process data points received from different sources (e.g. aggregate them as median, average or mode over a certain period of time)
- the format in which data will be delivered to customers in target blockchains

Each operator of a Gravity node can develop and use custom implementations according to the described specification. The development of the specification, documentation, or implementation of extractors are managed as part of the open-source development flow by Gravity developers.

Extractor services are managed and run independently on the servers of Gravity node operators. For operators, the possibilities of integrating extractors in the form of plugins, implementing custom versions of extractors, or supporting certain types of extractors, are optional and opt-in.

Gravity nodes communicate with each other through the account communicator module by sending transactions in the internal distributed ledger. The internal ledger is also used to read logs of messages from other nodes.

*Smart Contracts and Accounts Structure*

In addition to the network of nodes, the protocol incorporates infrastructure deployed within target blockchains, where information from data feeds is transmitted to. Figure 3 shows the components of the blockchain architecture implementing the Gravity protocol.

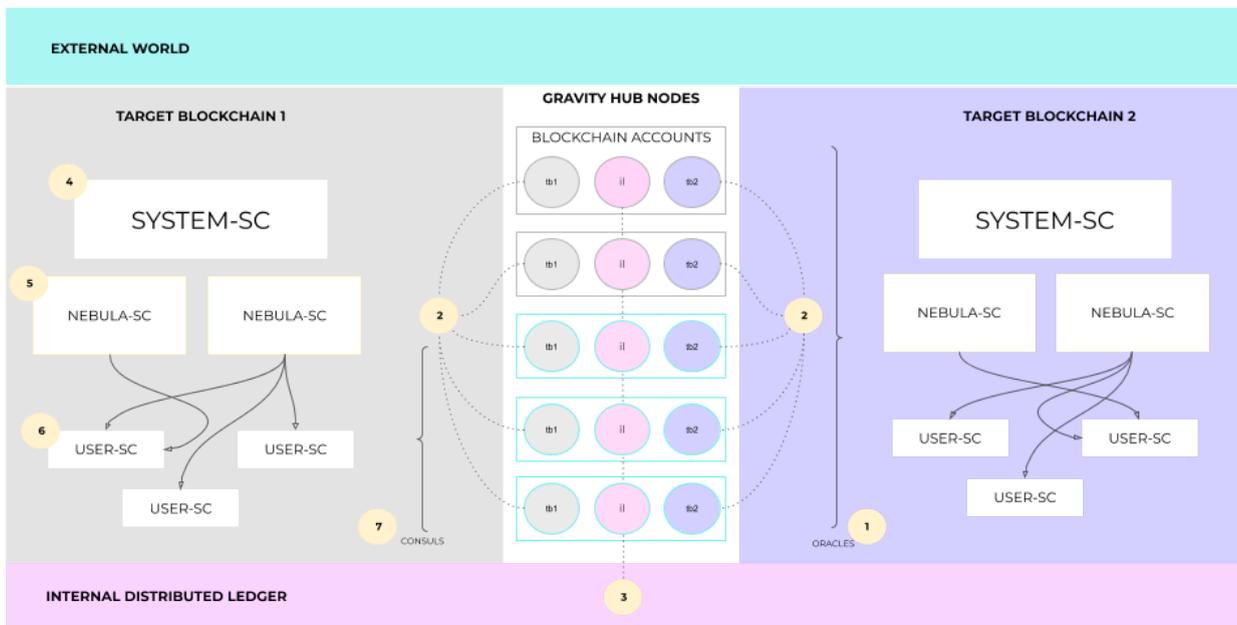

Figure 3: Smart contract and structure:

1) The Gravity system represents a community of nodes, where each node is an oracle for a specific target chain;

2) Each node has accounts in each supported target chain and in the internal ledger;

3) The Internal Distributed Ledger with a fast consensus finalization (e.g.: pBFT) enables communication between Gravity nodes, commit-reveals, aggregation, accounting and calculation of p2p scores;

4) The system smart contract in a given target chain serves as a register of all Gravity nodes and their scores, and a register of supported data feeds. It also stores and manages node deposits;

5) The Nebula smart contract, which is used to verify and access data from oracles, is the register of user subscriptions and the account that collects payments and distributes them among oracles;

6) The User smart contract contains subscriptions for data delivery from Gravity; A single user contract can receive data from different data feeds;

7) Consuls are a selection of several oracles with maximum scores in the network. They are authorized to update/migrate the SYSTEM-SC and serve as consensus validators in the internal ledger;

The described system is sufficiently flexible to manage multiple connected data feeds and subscriptions. This multi-contract approach allows for simplicity for implementation, increased security, and easy code audit, including formal verification of individual components.

## Reputation Management

The key feature of Gravity binding it together and giving it integrity as a system or as a decentralized service that can be trusted, is the protocol of participants' reputation management. It is due to this decentralized reputation system that such properties as self-management in Gravity networks are achieved, along with secure mechanisms of disconnecting malicious nodes and protection against attacks on the network through the production of malicious nodes, which may suddenly constitute the majority. In addition, the flexibility of Gravity reputation management is also an important advantage that allows attracting and retaining new participants, with a high reputation in the industry, allowing the expansion of the user base of the operating Gravity network.

The internal ledger, a software bus that nodes use to communicate with each other, processes nodes' trust ratings of their peers and aggregated reputation values. The Gravity nodes evaluate each others' performance and send their own confidence score for all other nodes of Gravity, which is then translated into the Gravity score for a particular node via the EigenTrust algorithm [6]. Figure 4 shows an example of pairwise estimates of four nodes.

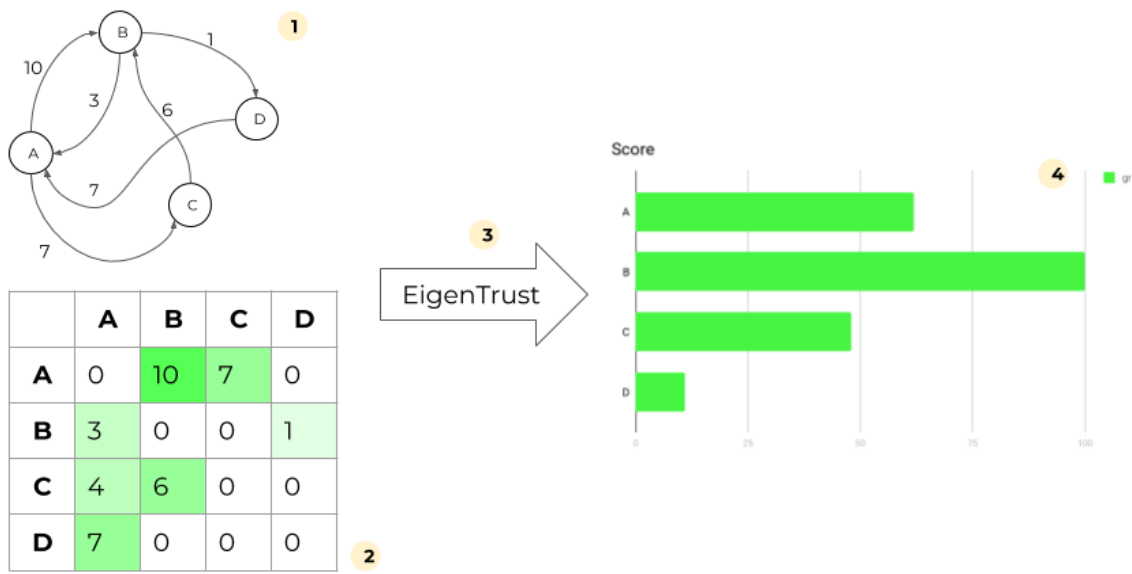

Figure 4: Reputation management based on p2p scores:

1) All operational nodes assess their trust level to other nodes of Gravity in an eventful and asynchronous manner;

2) The matrix of paired trust estimates is formed;

3) The matrix of estimates is transformed into a vector of Gravity scores, using the EigenTrust algorithm;

4) The obtained score values allow for a sorted ranking of all Gravity nodes by their scores normalized from 0 to 100.

All peer estimates are formed independently, based on the activity history of the evaluated node in the Gravity network, as well as on any relevant information from the outside world.

Peer scoring is based on:
1) automatic scoring:
    a) sending automatic shutdown signals:
        i) if the data from the oracle diverges greatly from the aggregate values,
        ii) if the node stops receiving and processing data feeds from its peers,
        iii) if the node stops responding to any requests from the Gravity service.

In this case, the peers can set their rating to zero, which guarantees a quick shutdown of the node to protect the Gravity network from malicious nodes.

   b) A periodic gradual Gravity score build-up for new nodes, based on the stability of operation for the duration of predefined time periods.

After reaching a certain value of the Gravity score, the node becomes able to participate in NEBULA-SC data provision for the target blockchains. Initially, a new node gets assigned a zero score, which is a defence measure against potential network capture attacks by a collusion of new malicious nodes, and also leaves time for manual intervention described below.

2) A manual scoring by node operators. The manual scoring procedures are the key element that underlies the governance mechanism of the Gravity network. Via manual scoring, the process of score modification takes place either for the nodes disconnected as a result of automatic shutdown, as described in 1.1 after an incident, or when it is necessary to manually "score-boost" nodes of the operators with an established reputation in the industry (for example, a large exchange or a popular data aggregator). As soon as the manual estimation occurs, the node that has received a grade ceases to collect automatic scores from the peer who initiated the manual rating procedure. For example, when the owners of Node A, which suspect Node B in some malicious behaviour, manually rank Node B down to decrease its score, Node A ceases to send evaluations about Node B, whereas a separate Node C will nevertheless continue the evaluation of Node B unless it also engages in a manual evaluation.

Ratings are recalculated on an event-driven basis, except for 1.1, where the recalculation is periodic.

Together, the collected assessments of peers make up a NxN table data structure, where N is the total number of Gravity nodes. Certain transformations of such matrices, conducted according to the EigenTrust algorithm, allow for the calculation of the final score for each node. The nodes constantly update their estimates of the adjacent nodes in the internal distributed ledger, which is used by the consuls as the source of Gravity score estimates when writing into a target blockchain. Thus, the consistency of estimations between all of the used target chains is achieved.

contentPupyshev et al. | Gravity: a blockchain-agnostic cross-chain communication and data oracles protocol | A preprint*EigenTrust Algorithm Pseudocode*

$$\vec{t}^{(0)} = \vec{p};$$
**repeat**
$$\quad \vec{t}^{(k+1)} = C^T \vec{t}^{(k)};$$
$$\quad \vec{t}^{(k+1)} = (1-a)\vec{t}^{(k+1)} + a\vec{p};$$
$$\quad \delta = ||t^{(k+1)} - t^{(k)}||;$$
**until** $\delta < \epsilon;$

where C is a matrix of $[c_{ij}]$ − normalized local trust value, $c_{ij} = \dfrac{max(s_{ij}, 0)}{\sum_j max(s_{ij}, 0)}$ ; $\vec{t}$ − global trust vector

Further details are described in the paper by Sepandar Kamvar and Mario Schlosser [6].

## Network Setup

Several nodes with the highest score become consuls. The consuls form the consensus core of the internal ledger, such as the pBFT consensus to finalize blocks and transactions. The composition of the consul committee is not fixed and depends on the score of participants.

A SYSTEM-SC is deployed in each of the supported blockchain networks, which takes care of system interactions and deposit locking, registration of nodes and collection of Gravity node scores about each other. A data provider can enter the network by locking a deposit for one year in any of the supported public host-chains in one of the tokens issued on the selected blockchain.

When entering the network, each new member is automatically assigned a default peer score value (evaluation from other Gravity nodes) of 0. Despite the fact that a Gravity node immediately becomes active after registration, it will be capable of participating in data delivery from extractors to target blockchains only when its reputation score reaches a certain level.

The exit from the Gravity network is free of charge. To leave the system, the node operator needs to call the deactivate method in SYSTEM-SC of the same blockchain where the registration took place. If at the moment of exit the reputation level is zero, the deposit is locked and can be released one year after exit, otherwise one year after the registration.

Locking a deposit above the threshold value produces no direct impact on the node's initial characteristics. However, a large deposit can serve as proof of the operator's intentions to work for the benefit of the entire network, which the node community may consider as a signal to manually increase the reputation ratings for their nodes.

# Pulse Consensus Algorithm

The Pulse algorithm is a two-step procedure that encompasses the core internal mechanics of Gravity as a data service, as well as the external parts of on-chain data verification and delivery within one of the supported target blockchains. Pulse is a process that starts with a request to deliver data to a target blockchain and ends with a successful verification and delivery of data to subscribers (USER-SC).

*Commit-Reveal & Data Aggregation Flow*

Figure 5 shows, in order, the main stages of data exchange, validation and aggregation for a particular data feed.

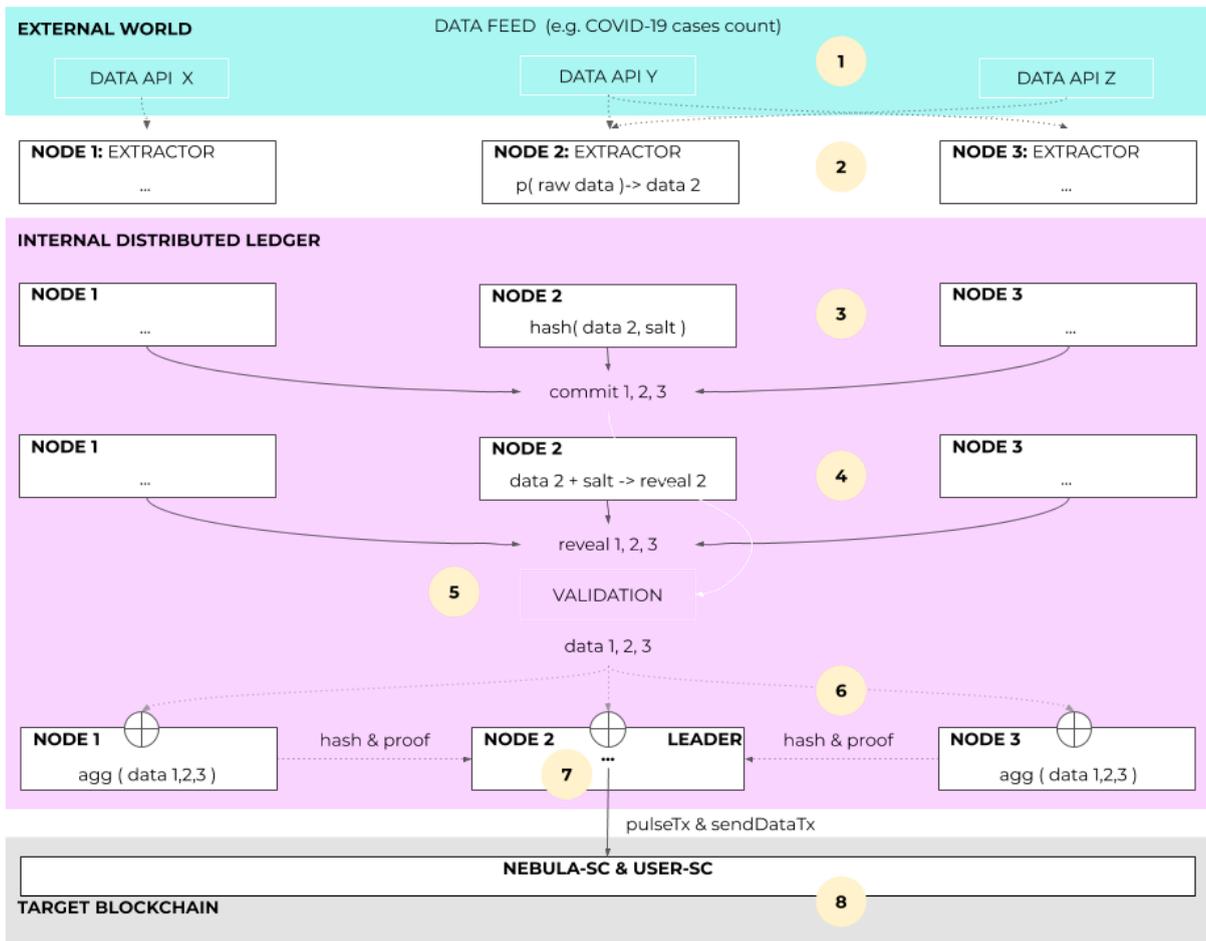

Figure 5: Pulse: Commit-reveal & data aggregation flow

1) Gravity nodes collect information from external sources. Several reliable sources can be used for one data feed;

2) The Node Extractor Module can collect data from a single source (Node 1) or accumulate raw data from multiple sources (Node 2, 3);

3) At the commit stage, each node generates a hash from data and writes it to the internal ledger;

4) At the reveal stage, the nodes disclose data and salt, for the purpose of checking it against the hash, and validate the fact that the data is taken from correct external sources. The reveal stage starts when at least K commits are collected, where K is defined by the rule of a sufficient number of signatures (e.g. the BFT rule, with 8 out of 11 commits available);

5) All validated data from oracles are available in the internal ledger for on-request reading by peers;

6) Each node aggregates its neighbours' data (e.g.: calculates the median).

7) Each node hashes the aggregated value and forms a cryptographic signature for it. All hashes and signatures are collected by an oracle which is the Leader of the pulse;

8) The Leader performs several transactions in the target chain: pulse tx for threshold signature verification and sendData tx for data transfer to user smart contracts.

*Threshold Signature & Data Verification Flow*

Figure 6 presents, in order, the main stages of generating a hash, a cryptographic proof for data transmitted from each node, and verifying the threshold signature conditions.

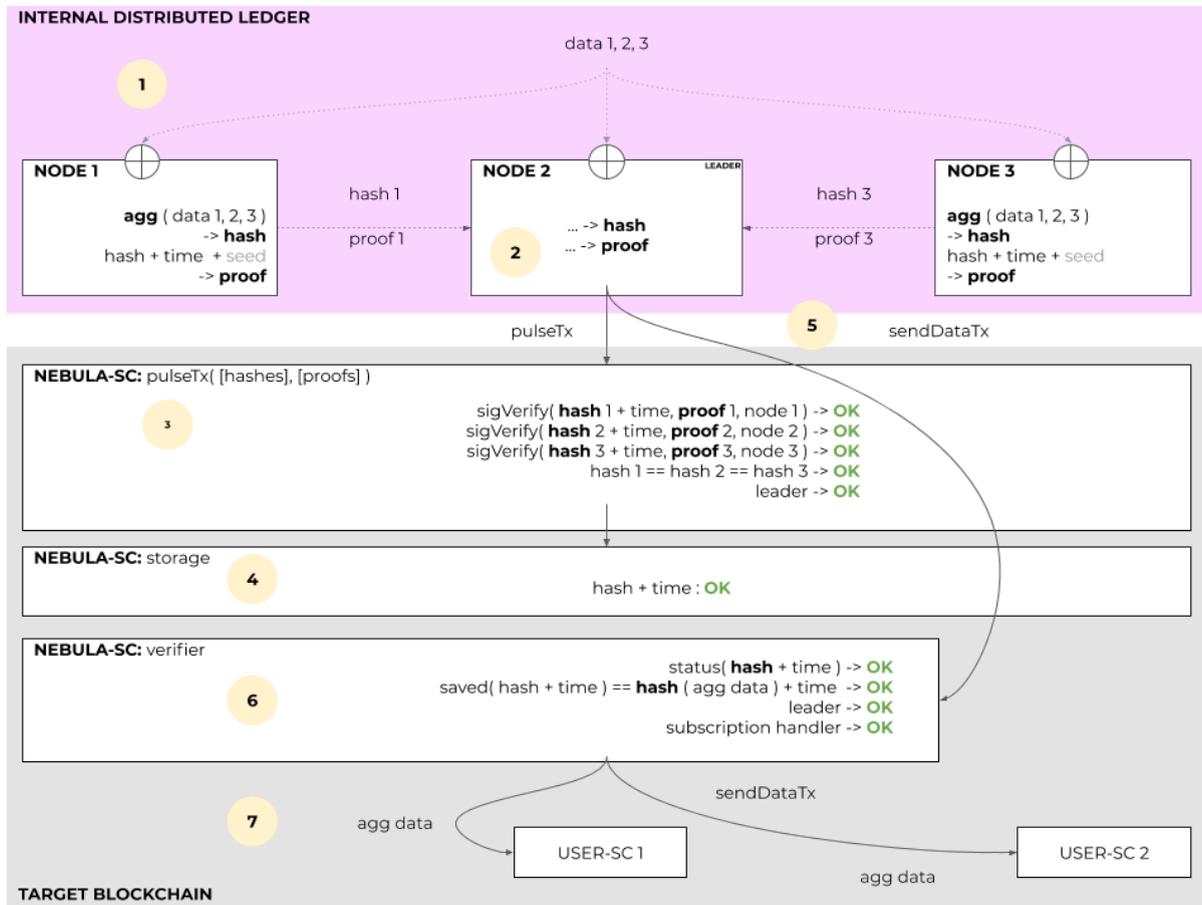

Figure 6: Pulse: Threshold signature & data verification flow

1) Each node aggregates other nodes' data (e.g.: calculates the median), hashes the aggregated value and generates a cryptographic signature for the hash using the current timestamp and the data feed id;

2) The Pulse Leader collects all hashes and their signatures from the peers, reading them from the internal ledger. The Pulse Leader performs a pulse transaction in the target chain by passing the hashes and signatures to the corresponding nebula contract;

3) Inside the contract, within the execution of pulse tx, the correspondence of signatures to signatories is verified, with the actual timestamp, the nebula, and the leader. At least K hashes should be valid, where K is defined by the rule of a sufficient number of signatures (this parameter is set in the NEBULA-SC, e.g. 8 out of 11);

4) The Pulse status (for the data hash, timestamp and other verifiable parameters) is written into the storage of the nebula contract;

5) The Pulse Leader forms a transaction on behalf of NEBULA-SC, which supplies data to USER-SC;

6) As part of the outgoing transaction verification, the pulse verification status, the validity of hashes from the aggregated data and the data transmitted to the user contract, and the recipient itself, are verified;

7) Within one verified data pulse, the leader calls user methods which accept the verified data.

\* The Nebula id is a specifier for the contract, the receiving port of the data feed. It is needed for verification of the signature intended for a particular nebula contract. For the sake of simplicity, some stages are omitted in the illustration.

Users can subscribe their USER-SCs to Gravity events, which will result in specified functions being called with aggregated data sent as parameters into the functions. Depending on the subscription configuration, users can change security settings, adjust pricing, or set up white lists of oracles.

*Pulse Consensus Pseudocode*

**Begin**
> **Each node:** Check if $O_i$ is a provider selected from $\{O_i\}_N$
> **Each node:** Send request to scheduler for `params(t)`
> **Each node:** Send request to extractor with `params(t)`
> **Each node:** Process raw data
> **Each node:** Broadcast commits to peers $\{O_i\}_N$.
>> `commit = hash(reveal); reveal = message+proof`
>
> **IDL**: Wait until at least K (threshold signature rule) commit messages from distinct $O_i$ are received, where `K<=N`
> **Each node:** Broadcast reveal messages to the peers $\{O_i\}_K$
> **IDL**: Verify all received commits and reveals, check the threshold signature rule (i.e. 8 out of 11)
> **Each node:** Compute agg. data from reveals of the peers $\{O_i\}_K$
> **Each node:** Execute `hash(agg_data)` and generate proof from hash, timestamp & seed
> **Leader**: Call `pulse-tx` invocation with peer's hashes & proofs in nebula contract
> **Nebula**: Verify hashes, timestamp, proofs, threshold signature rule, and leader
> **Leader**: For all subscribers $s_j$ from $\{s_j\}_S$,
>> call `sendDataTx` with agg. data verified by nebula contract

**End**

# Economy

*Decentralized Treasury*

Gravity protocol is designed as an open source blockchain-agnostic protocol without its own utility token. The options available for the support of long-term financial sustainability, as well as for sponsoring the development and research by different teams and attracting interested sponsors, are constrained in such systems and require an innovative approach. To solve this task, we are proposing a model of taxation/fee for each transaction to register at Gravity services or payment for data provision from USER-SC. All funds collected under this model go to a decentralized cross-chain treasury fund, where they can be managed by a special decentralized autonomous organization (DAO).

The activity of the DAO will be directed towards achieving long-term financial sustainability of the protocol development. The treasury fund may be replenished by any external project sponsors interested in developing and promoting the protocol. The description of architecture and mechanisms of DAO functioning is beyond the scope of this paper and will be outlined in new documentation dedicated to this model.

## Network Monetisation

Gravity users can choose to use one of two options to pay for Gravity services: putting a deposit on a designated blockchain account and maintaining it above a certain threshold, or paying a recurring subscription fee.

In the first case, the users can deposit tokens by calling the deposit method in NEBULA-SC. As long as their balance is sufficient, the USER-SC will receive messages with data from Gravity oracles.

In the second case, the payment takes part alongside each execution of a user contract method (USER-SC) and is sent to the NEBULA-SC account. If payment is not received during this operation, the subscription is considered to be suspended and requires reactivation by the user.

All undistributed profits are accumulated on Gravity smart contracts and distributed every week. Gravity smart contracts (SYSTEM-SC and NEBULA-SC) keep activity logs for each oracle in all host blockchains. At the moment of profit distribution, the percentage of funds that can be claimed by each of the Gravity nodes is calculated, where the share of rewards received by a single oracle is the product of its activity and reputation score normalized on the total amount of funds.

In each of the contracts that the oracle participated in, it has the ability to withdraw funds proportional to its impact share.

The SYSTEM-SC is replenished from a pool filled by NEBULA-SC created on the respective host blockchains. NEBULA-SC contracts receive payments with each operation of data delivery to a USER-SC.

Example

The NEBULA-SC that handles the data feed of "COVID-19 cases" has collected 100 TOKENS from subscribers during the week. Node V has been supplying 10% of all verified data hashes. At the moment of calculation, the Gravity Score for the node is 85gr. Thus, the unnormalized value of impact on the nebula's performance is $0.1*0.85 = 0.085$. Let us assume that the remaining six nodes have impact values [ 0.2*0.7, 0.2*0.7, 0.2*0.7, 0.1*0.6, 0.0*0.9 ]. The sum of all impacts is equal to 0.705. This means that during the last week of operation, Node V produced a normalized impact equal to around 12%. As a result, Node V can collect 12 tokens from the weekly undistributed profit of NEBULA-SC.

# Applications

*Data provision*

Data provision capabilities of the protocol are well covered in this document's main part, but it is especially important to note that the data provision scheme is the foundation for higher-level and specialized protocol applications described below.

*Triggers*

While the protocol described in this paper is focused on delivering network-agreed oracle data to target blockchains, another worthwhile functionality from a practical standpoint is to regularly call specified methods of user smart contracts without transmitting any data to them. For instance, the invocation of a smart contract function can be tied to the true-false outcome of an event in a blockchain or on a public web portal. In this case, the Gravity system would be capable of reporting when the event occurred. These applications are called triggers and can be implemented on the Gravity protocol with minimal modifications and customizations of NEBULA-SC and USER-SC.

A similar flow can be implemented for so-called pacemaker oracles, which are used as external event triggers for those smart contract systems where contracts are simply verifier scripts without loops and recursions and may not be Turing-complete [7].

*Cross-chain Transfer Gateways*

Another useful application for the Gravity protocol is the cross-chain transfer of digital assets (tokens). Data provision between two blockchain networks is the key link for such applications, with the primary principle being to lock tokens in one blockchain (origin) and to report this event to another blockchain (destination), which will issue the exact same number of tokens. In the case of an emergency situation where the issuance of the token has not occurred, a dispute resolution is started, which sends a signal to unlock the token on the origin blockchain. This approach to gateways requires a coordinated operation between oracles of the Gravity network in both target blockchains. There exist multiple approaches to implement such applications, for example, based on Merkle trees, if supported by the available cryptography.

A different approach to implementing gateways is creating a signal provider from specialized data feeds with the prices of a particular asset. Building gateway protocols comes down to designing an application layer protocol over the Gravity protocol. Notably, these protocols themselves can be fully decentralized and be based on smart, actorless contracts to avoid introducing centralized points of failure into the system.

*Cross-chain Delegation Gateways*

Similarly to cross-chain transfers, the same approach is applicable to other practical data solutions. With cross-chain delegation gateways, an account holder in one blockchain (origin) is able to create events in another blockchain (destination), meaning that a user of an application deployed in blockchain A can manage it directly from blockchain B without having an account in blockchain A. Implementing such applications requires a fairly complex logic of cross-chain signalling and stable infrastructure components including triggers and cross-chain transfers. We are confident that Gravity protocol can provide a solid foundation for the creation and promotion of cutting-edge cross-chain decentralized applications of this type.

*Sidechains*

A special case of cross-chain communication, which allows for scalability of an already deployed public network, is called sidechain technology. Typically, in order to confer economic value to the token of a sidechain, cross-chain providers should ensure that a certain amount of native tokens is locked on the origin blockchain and its corresponding total supply is locked on the sidechain. In order to manage these processes and to verify system changes in the sidechain or to react correctly to changes in the origin blockchain, a decentralized oracle is needed that provides signals through structurally related data feeds from one blockchain to another. The Gravity network is an example of such an oracle, which in combination with additional modules in the form of smart contracts in origin and sidechain blockchains allows for increased computation and transaction performance due to being spanned across two different blockchains but utilizing only one utility token from the origin blockchain.

# Summary

The concept of Gravity described in this paper allows for the implementation of a complex interconnected network of oracles, supporting communication of blockchain networks with the outside world, cross-chain communication and transfers, as well as integrating sidechains, within one holistic and self-governing system. Due to the proposed consensus design, Gravity can be regarded as a singular decentralized blockchain-agnostic oracle.

Pupyshev et al. | Gravity: a blockchain-agnostic cross-chain communication and data oracles protocol | A preprint# References

[1] Waves Platform. Gravity Hub to unite blockchains and connect with other systems. 2019. https://medium.com/wavesprotocol/gravity-hub-to-unite-blockchains-and-connect-with-other-systems-9e2f2a2818f

[2] Neutrino Protocol. Neutrino Protocol 2020: Vision & Challenges. 2020. https://medium.com/@neutrinoteam/neutrino-protocol-2020-vision-challenges-1fec59e1577b

[3] Jae Kwon & Ethan Buchman. Cosmos Whitepaper: A Network of Distributed Ledgers. 2019. https://cosmos.network/resources/whitepaper

[4] Steve Ellis & Ari Juels & Sergey Nazarov. ChainLink: A Decentralized Oracle Network. 2017. https://www.allcryptowhitepapers.com/wp-content/uploads/2018/05/ChainLink.pdf

[5] NEM: Technical Reference. 2018. https://nem.io/wp-content/themes/nem/files/NEM_techRef.pdf

[6] Sepandar Kamvar & Mario Schlosser & Hector Garcia-Molina. The EigenTrust Algorithm for Reputation Management in P2P Networks. 2003. https://www.researchgate.net/publication/2904367_The_EigenTrust_Algorithm_for_Reputation_Management_in_P2P_Networks

[7] Marc Jansen & Farouk Hdhili & Ramy Gouiaa & Ziyaad Qasem. Do Smart Contract Languages Need to be Turing Complete? 2019. https://www.researchgate.net/publication/332072371_Do_Smart_Contract_Languages_Need_to_be_Turing_Complete

[8] Waves Platform. Ecosystem development strategy: interchain bridges. 2019. https://medium.com/wavesprotocol/ecosystem-development-strategy-interchain-bridges-8b09e9f48539

[9] Tellor Core. The current oracle landscape. 2020. https://medium.com/@tellor/the-current-oracle-landscape-f772356694e2

[10] Provable Things. Authenticity proofs verification: off-chain vs on-chain. 2017. https://medium.com/provable/authenticity-proofs-verification-off-chain-vs-on-chain-64a95fee097b

[11] Bitmex Research. Ethereum 2.0 tech overview. 2020. https://blog.bitmex.com/ethereum-2-0/

[12] Amritraj Singh & Kelly Click & Reza M. Parizi. Sidechain technologies in blockchain networks: An examination and state-of-the-art review. 2019. https://www.researchgate.net/publication/336772695_Sidechain_technologies_in_blockchain_networks_An_examination_and_state-of-the-art_review21